# Anomalous Size Dependence of the Thermal Conductivity of Graphene Ribbons

Denis L. Nika[1,2], Artur S. Askerov[2] and Alexander A. Balandin[1*]

[1]Nano-Device Laboratory, Department of Electrical Engineering and Materials Science and Engineering Program, Bourns College of Engineering, University of California, Riverside, CA 92521 U.S.A.

[2]E. Pokatilov Laboratory of Physics and Engineering of Nanomaterials, Department of Theoretical Physics, Moldova State University, Chisinau, MD-2009, Republic of Moldova

---

[*] Corresponding author; electronic address: balandin@ee.ucr.edu ; NDL web-site: http://ndl.ee.ucr.edu

Context Image

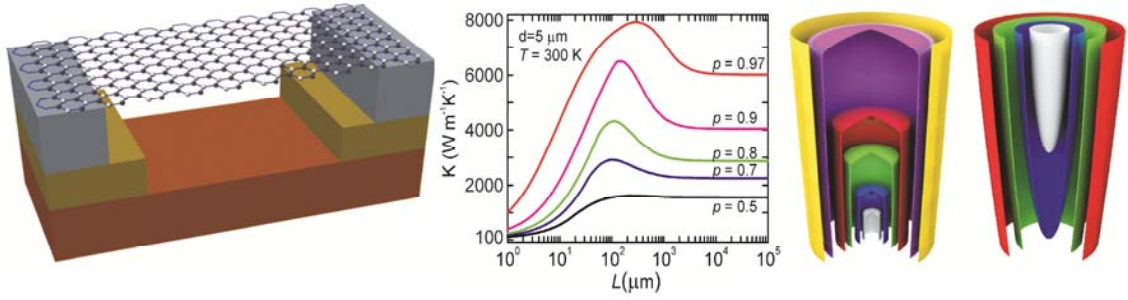

**Abstract**

We investigated the thermal conductivity $K$ of graphene ribbons and graphite slabs as the function of their lateral dimensions. Our theoretical model considered the anharmonic three-phonon processes to the *second-order* and included the *angle-dependent* phonon scattering from the ribbon edges. It was found that the long mean free path of the long-wavelength acoustic phonons in graphene can lead to an unusual *non-monotonic* dependence of the thermal conductivity on the length $L$ of a ribbon. The effect is pronounced for the ribbons with the smooth edges (specularity parameter $p>0.5$). Our results also suggest that – contrary to what was previously thought – the bulk-like 3D phonons in graphite can make a rather substantial contribution to its in-plane thermal conductivity. The Umklapp-limited thermal conductivity of graphite slabs scales, for $L$ below ~ 10 μm, as $\log(L)$ while for larger $L$, the thermal conductivity approaches a finite value following the dependence $K_0 - A \times L^{-1/2}$, where $K_0$ and $A$ are parameters independent of the length. Our theoretical results clarify the scaling of the phonon thermal conductivity with the lateral sizes in graphene and graphite. The revealed anomalous dependence $K(L)$ for the micrometer-size graphene ribbons can account for some of the discrepancy in reported experimental data for graphene.

Thermal transport in two-dimensional (2D) and one-dimensional (1D) material systems attracts increasing attention owing to the fundamental nature of questions and practical importance of the subject [1]. Theoretical predictions that the *intrinsic* thermal conductivity $K$ – limited by the crystal anharmonicy alone – can diverge with the crystal size $L$ in 2D and 1D systems, continue to ignite debates [1-16]. Theoretical studies of the lattice thermal transport in 2D anharmonic Fermi-Pasta-Ulam (FPU) lattices [2-3], 2D anharmonic Lennard-Jones lattices [3], 2D harmonic lattices with disorder [4] and 1D FPU chains [2-3, 5] suggested that the lattice, i.e. phonon, thermal conductivity $K$ diverges as $\log(L)$ for 2D lattices and as $L^\alpha$ for 1D chains ($\alpha<1$), where the length $L$ is proportional to the number of the lattice points $N$ along the heat propagation direction [2-5].

Examples of the studies that revealed the divergence in 2D thermal conductivity are numerous. Wang and Lee [6] concluded that in 1D chain, with both longitudinal and transverse motions of atoms, the thermal conductivity diverges as $\sim\log(N)$ or $N^\alpha$, depending on the strength of transverse interactions. For the strong transverse restoring force, $K$ diverges as $\log(N)$, for the intermediate strength – as $K\sim N^{1/3}$, and for the weak strength – as $K\sim N^{2/5}$ [6]. Dimensional crossover of the thermal conductivity, in the FPU lattices, was studied computationally depending on the parameter $\delta=N_x/N_y$, where $N_x$ ($N_y$) is the number of the lattice sites along the longitudinal (transverse) direction $x$ ($y$) [2]. It was found that $K\sim N^\alpha$ for $\delta\ll 1$ (1D case) while $K\sim\log(N)$ otherwise (2D case).

From the other side, there were studies that suggested that Fourier's empirical law of thermal conduction is valid for 1D and 2D systems [7-9]. Casati et al. [7] numerically found the finite value for the 1D many-body chaotic system of $N$ particles. Jackson and Mistriotis [8] investigated 1D and 2D lattices and found that there exist a range of the lattice parameters for $K$ transition infinite to finite value. Numerous molecular dynamic (MD) simulations have given contradictory results [10-12]. Yao et al. [10] and Zhang and Li [11] concluded that the lattice thermal conductivity diverges in a carbon nanotube (CNT) with increasing length analogously to the strictly 1D systems. At the same time,

Donadio and Galli [12] demonstrated the non-divergent thermal conductivity in CNTs using both MD and the Boltzmann transport equation (BTE) approach.

A notable study by Mingo and Broido [13] established that the non-divergent thermal conductivity in CNTs results from the three-phonon anharmonic processes of the second or higher order. If the three-phonon processes are considered to the second order, the thermal conductivity initially increases with CNTs length but then saturates to some well-defined finite value. This result indicates that an accurate treatment of the intrinsic thermal conductivity of the low-dimensional systems should include the three-phonon anharmonic processes of the higher order, i.e. beyond the conventional first-order Umklapp scattering. Addition of the crystal lattice disorder or diffuse interface scattering can eliminate the need for the second-order three-phonon scattering in obtaining the finite thermal conductivity for the low-dimensional systems [5, 14]. Most studies agree that in three-dimensional (3D) crystals the intrinsic thermal conductivity – limited by the anharmonicity alone – has non-divergent value [4, 15]. The first-order three-phonon Umklapp scattering is sufficient for obtaining the finite $K$ in 3D [13].

Experimental studies of thermal transport in low-dimensional systems have been performed using CNTs [16-18] as quasi 1D system and graphene [19-25] as 2D system. In general very high values were reported for both CNTs and graphene [16-25]. Suspended few-layer graphene (FLG) flakes were utilized to study the crossover of thermal transport as the system dimensionality changed from the quasi-2D graphene to quasi-3D graphite films [21]. The strong length dependence was observed experimentally in CNTs, which prompted suggestions of the breakdown of Fourier's empirical law [16]. The latter is related to the intrinsic thermal conductivity divergence discussed above.

Although the strong size dependence was reported in many studies of heat conduction in graphene or CNTs [13,16-17,19-21,25-33], it is usually difficult to distinguish among the various possible mechanisms. Among them are the $K(L)$ dependence in the ballistic thermal transport regime where $L<<\Lambda$, the $K$ dependence on the nanoribbon width due to the acoustic phonon – rough edge scattering, or the fundamental $K$ size dependence in 1D or 2D lattices where anharmonic interactions are not sufficient for establishing finite $K$ over the given length scale $L$. These important questions call for a rigorous study of the lateral size effects on the thermal conductivity of graphene ribbons and graphite slabs. Consideration of the graphene and graphite together is needed in order to elucidate the differences in 2D and 3D phonon transport.

In this letter we report on the theoretical study of the thermal conductivity in graphene ribbons (see the geometry and notations in Figure 1), which takes into account the anharmonic three-phonon processes to the second-order together with the angle-dependent phonon scattering from the ribbon edges. The proper inclusion of the angle dependence to the edge scattering allowed us to reveal an unusual non-monotonic dependence of the thermal conductivity on the ribbon length. Owing to the exceptionally long phonon mean free path (MFP) in graphene, the abnormal $K(L)$ dependence can manifest itself in the ribbons of the tens-of-micrometers lengths. Revisiting the theory of thermal conduction in bulk graphite we found that contrary to all previous assumptions the bulk-like 3D phonons make substantial contribution to the thermal conductivity of graphite. The rest of the letter we first address the thermal conductivity in graphite slabs and then in graphene ribbons.

[Figure 1]

We start by revisiting the calculations of the thermal conductivity of graphite reported in the classical works of Klemens and co-workers [34-35]. In these papers, Klemens assumed that the phonon transport in graphite is essentially 2D-like for all phonon frequencies $\omega$ above a certain low-bound cut-off frequency $\omega_c$. The phonons with $\omega > \omega_c$ are referred to as 2D phonons while those with $\omega \leq \omega_c$ are called 3D phonons. The contribution of low-frequency, i.e. long wavelength, 3D phonons to the in-plane thermal conductivity of graphite was assumed to be negligible [34-35]. The physical reasoning was that these phonons will experience stronger scattering due to the inter-layer coupling and will not have long MFP. We re-examine this point in order to be able to provide a meaningful comparison with the free-standing graphene.

In order to determine the role of 3D phonons in the in-plane thermal conductivity one needs to start with an accurate phonon spectrum in graphite. We calculate it using the valence-force field (VFF) model of lattice dynamics. The details of the VFF calculations for graphene and few-layer graphene (FLG) were reported by us elsewhere [21, 26]. The phonon frequencies $\omega_s(\vec{q})$ of graphite calculated along Γ-A and Γ-M crystallographic directions are presented in Figure 2 for all phonon branches *s*. These polarization branches include (i) out-of-plane optical (*s*= ZO, ZO') and out-of-plane acoustic (*s*=ZA, ZA') phonons with the displacement vector normal to the basal planes; (ii) transverse optical (*s*=TO, TO') and transverse acoustic (*s*=TA, TA') phonons, which corresponds to the transverse vibrations within the basal plane; (iii) longitudinal optic (*s*=LO, LO') and longitudinal acoustic (*s*=LA, LA'), which corresponds to the longitudinal vibrations within the basal plane.

[Figure 2]

In Figure 3 we show the equal energy surfaces (EESs) $\omega_s(\vec{q}) = \omega_{const}$ in graphite for the *LA* (a) and *TA* (b) phonon branches. The surfaces are plotted for different values of $\omega_{const}$ from the range $0 \leq \omega_{const} \leq \omega_{s,max}^{\|}(M)$, where $\omega_{s,max}^{\|}(M)$ is the frequency of the phonon branch *s = LA, TA* at the *M* point of the Brillouin zone (BZ). Each EES has the shape close to cylindrical with the top and bottom covers for $\omega_{const} < \omega_{c,s}$, where $\omega_{c,s} \equiv \omega_s(A)$ is the frequency of the phonon branch *s* at the *A* point of BZ. The increase in $\omega_{const}$ leads to cylinder extension along $q_z$-axis, i.e. perpendicular to the basal plane of graphite. For almost all frequencies $\omega_{const} \geq \omega_{c,s}$, EESs have cylindrical-like shapes, which narrow for $q_z$ close to $q_{z,max}$. The narrowing reflects an increase in the out-of-plane vibration energy with increasing $q_z$ and corresponding reduction of the in-plane vibration energy and radius $q_\| = (q_x^2 + q_y^2)^{1/2}$ of the intersection of the cylinder and the plane $q=q_z$. Both the cylinder height $q_z$ and radius $q_\|$ increase with $\omega_{const}$ and, for $\omega_{const}$ between the frequencies of the phonon branch *s* at *M* and *K* points of BZ, the cylinders transform to the right-angle prisms or parts of the right-angle prisms with the hexagonal base (not shown in Figure 3). The latter originates from the hexagonal symmetry of the graphite unit cell.

[Figure 3]

The Umklapp-limited phonon thermal conductivity tensor for the basal plane of graphite can be written as [26]

$$K_{\alpha\beta} = \frac{1}{L_x L_y L_z} \sum_{s,\vec{q}} \hbar \omega_s(\vec{q}) \tau_{U,s}(\omega_s(\vec{q})) \upsilon_{\alpha,s} \upsilon_{\beta,s} \frac{\partial N_0}{\partial T}, \qquad (1)$$

where $\hbar$ is the Plank's constant, $L_x$, $L_y$, $L_z$ are the sizes of the graphite slab, $\tau_{U,s}$ is the phonon Umklapp scattering rate for $s$th phonon branch, $\omega_s$ is the phonon energy of $s$th phonon branch and $\upsilon_{\alpha,s}(\upsilon_{\beta,s})$ is the projection of group velocity of $s$th phonon branch on the axis $\alpha(\beta)$, $\vec{q}(\vec{q}_\parallel, q_z)$ is the three-dimension phonon wave vector, $N_0 = 1/(\exp[\hbar \omega_s / k_B T] - 1)$ is the Bose-Einstein distribution function, $T$ is the absolute temperature, $k_B$ is the Boltzmann's constant and $\varphi$ is the angle between the in-plane phonon wave vector $\vec{q}_\parallel$ and the temperature gradient.

To better elucidate the dependence of the in-plane thermal conductivity of graphite on the sample lateral dimensions we derive an analytical expression by approximating EESs of graphite (see Figure 3) with the cylindrical surfaces $\omega_s(\vec{q}) = \omega_s^\parallel(\vec{q}_\parallel)\theta(|q_{z,c}| - |q_z|)$, where $\theta(|q_{z,c}| - |q_z|) = 1$ if $|q_{z,c}| - |q_z| > 0$ and $\theta(|q_{z,c}| - |q_z|) = 0$ otherwise. The value of $q_{z,c}$ is determined by $\omega_s^\parallel$: $q_{z,c} = \omega_s^\parallel / \upsilon_s^\perp$, where $\upsilon_s^\perp = \omega_{c,s} / q_{z,\max}$. The summation in Eq. (1) is performed over longitudinal (*LA*) and transverse (*TA, ZA*) phonon branches. The total thermal conductivity of graphite is given by: $K \equiv K_{xx} = \sum_{s=LA,TA,ZA} K_s^{\omega \leq \omega_{c,s}} + K_s^{\omega > \omega_{c,s}}$, where $K_s^{\omega \le \omega_{c,s}}$ and $K_s^{\omega > \omega_{c,s}}$ are the contributions to the thermal conductivity from inner (3D phonons) and outer (2D phonons) area of EES $\omega_s(\vec{q}) = \omega_{c,s}$, respectively:

$$K_s^{\omega \le \omega_{c,s}} = \frac{1}{8\pi^3} \int_{\omega_{\min,s}}^{\omega_{c,s}} \int_0^{2\pi} \hbar \omega_s^{\parallel}(q_{\parallel}) \tau_{U,s}(\omega_s^{\parallel}) v_s^{\parallel}(q_{\parallel})) \cos^2 \varphi \frac{\partial N_0}{\partial T} q_{\parallel} d\omega_s^{\parallel} d\varphi \int_{-q_{z,c}}^{q_{z,c}} dq_z =$$

$$= \frac{\hbar^2}{4\pi^2 v_s^{\perp} k_B T^2} \int_{\omega_{\min,s}}^{\omega_{c,s}} [\omega_s^{\parallel}(\vec{q}_{\parallel})]^3 \tau_{U,s}(\omega_s^{\parallel}) v_s^{\parallel}(q_{\parallel})) \frac{\exp(\hbar \omega_s^{\parallel}/k_B T)}{[\exp(\hbar \omega_s^{\parallel}/k_B T) - 1]^2} q_{\parallel} d\omega_s^{\parallel}. \quad (2)$$

$$K_s^{\omega > \omega_{c,s}} = \frac{1}{8\pi^3} \int_{\omega_{c,s}}^{\omega_{\max,s}} \int_0^{2\pi} \hbar \omega_s^{\parallel}(q_{\parallel}) \tau_{U,s}(\omega_s^{\parallel}) v_s^{\parallel}(q_{\parallel}) \cos^2 \varphi \frac{\partial N_0}{\partial T} q_{\parallel} d\omega_s^{\parallel} d\varphi \int_{-q_{z,\max}}^{q_{z,\max}} dq_z =$$

$$= \frac{\hbar^2 \omega_{c,s}}{4\pi^2 v_s^{\perp} k_B T^2} \int_{\omega_{c,s}}^{\omega_{\max,s}} [\omega_s^{\parallel}(q_{\parallel})]^2 \tau_{U,s}(\omega_s^{\parallel}) v_s^{\parallel}(q_{\parallel}) \frac{\exp(\hbar \omega_s^{\parallel}/k_B T)}{[\exp(\hbar \omega_s^{\parallel}/k_B T) - 1]^2} q_{\parallel} d\omega_s^{\parallel}. \quad (3)$$

Substituting the expression for the phonon Umklapp relaxation time $\tau_{U,s} = M v_s^2 \omega_{\max,s}/(\gamma_s^2 k_B T [\omega_s^{\parallel}]^2)$, we performed calculation of $K_s^{\omega \le \omega_c}$ and $K_s^{\omega > \omega_c}$ from Eqs. (2-3) using the actual phonon energies and group velocities in graphite (see Figure 2). In Eq. (2) the low-bound cut-off frequency $\omega_{\min,s}$ depends on the in-plane size $L$ of the graphite sample and is determined from the condition that the in-plane phonon MFP cannot exceed $L$ of the sample $L = \tau_{U,s}(\omega_{\min,s}) v_s$ [27, 30, 35], i.e. $\omega_{\min,s} = \sqrt{M[v_s^{\parallel}]^3 \omega_{\max,s}/(k_B T L \gamma_s^2)}$, where $M$ is the graphene unit cell mass and $\gamma_s$ is the branch-dependent average Gruneisen parameter. Note that for the infinitely large graphite sample $\omega_{\min,s} = 0$.

Several research groups have reported different values of the average Gruneisen parameters in graphite, ranging from $\gamma = 1$ to $\gamma = 2$ [34-38]. The fact that the Gruneisen parameter in graphene and graphite is a strong function of the phonon polarization branch is known [39]. For this reason, in our calculations, we use separate Gruneisen parameters for each phonon branch, obtained by averaging of a mode-dependent Gruneisen parameters over the relevant phonon wave-vector ranges: $\gamma_{LA} = 2$, $\gamma_{TA} = 1$ and $\gamma_{ZA} = -1.5$. With these parameters, the calculated RT thermal conductivity for infinite graphite slab ($L \to \infty$, $\omega_{\min,s} = 0$) is $K = 1900$ W/mK, which is in good agreement with experimental values for the highly-oriented pyrolytic graphite [40]. The unexpected finding, is that the contribution of the long-wavelength 3D phonons $K_s^{\omega \leq \omega_{c,s}}$ to the thermal conductivity, which was assumed as negligible in the earlier works [34-35], is large and constitutes ~ 50% for *LA* phonons and ~ 40% for *TA* phonons. We also checked the validity of the assumption that all phonon modes in graphite are populated at RT, which is often used in the thermal conductivity calculations. Our results show that in the case of graphite this assumption overestimates $K_s^{\omega > \omega_{c,s}}$, i.e. the contribution of 2D phonons, by a factor of ~1.45 for *LA* branch and ~1.15 for *TA* branch.

Figure 4 shows the dependence of the thermal conductivity of graphite on the slab length *L* for different temperatures. For small *L*, when $\omega_{\min,s} > \omega_{c,s}$, $K^{3D} = \sum_{s=LA,TA,ZA} K_s^{\omega \leq \omega_{c,s}} = 0$, while $K^{2D} = \sum_{s=LA,TA,ZA} K_s^{\omega > \omega_{c,s}} \sim \log L$, therefore total thermal conductivity scales as $K \sim \log L$. For larger *L*, when $\omega_{\min,s} \leq \omega_{c,s}$, $K^{2D}$ is independent on *L*, while $K^{3D} \sim (A - BL^{-1/2})$ and the total thermal conductivity increases with increasing *L* as $K = (K_0 - BL^{-1/2})$, where $K_0$, *A* and *B* are parameters independent of *L*. When the sample size $L \to \infty$, then $\omega_{\min,s} \to 0$ and the thermal conductivity approaches bulk graphite limit. For the realistically chosen material parameters, the thermal conductivity of graphite approaches the bulk limit at *L*~10 μm. The obtained results are unexpected in two accounts. First, the contribution of the 3D bulk phonons to the thermal conductivity of graphite along the basal planes is much larger than it was previously believed. Second, there is fundamental lateral size dependence of the thermal conductivity of bulk graphite all the way up to the length *L*~10 μm. This dependence is different from the size dependence in the ballistic transport regime. It manifests itself in graphite up to a rather large length scale *L* owing to the large phonon MFP in graphite basal planes. Our analytical derivations allowed us to explicitly reveal these phenomena.

[Figure 4]

Now we turn to thermal transport in graphene ribbons. The schematic view of a suspended graphene ribbon used in the experiments and pertinent notations are shown in Figure 1. Unlike in graphite, the phonon transport in graphene is two-dimensional all the way down to $\omega = 0$. The long-wavelength phonons weakly scatter in the three-phonon Umklapp processes in graphene calculated to the first order [13,26,28,34-35] resulting in divergent $K$. In his treatment of thermal conductivity of graphene, Klemens overcame the problem of the long-wavelength phonons by introducing the size-dependent cut-of frequency $\omega_{\min,s}$ defined by the equation $L = \tau(\omega_{\min,s}) \cdot v_s^{\parallel}$ [35]. This approach leads to the logarithmic dependence of the thermal conductivity on $L$, which is in line with the results obtained for the ideal 2D lattices [2-3]. However, the dependence $K \sim \log(L)$ is obtained using a number of simplifications, e.g. treatment of the anharmonic phonon scattering to the first order [27,30,35]. This model is not suitable for the large graphene samples when other scattering mechanisms, e.g. multi-phonon processes, scattering on edges, grains and crystal lattice imperfections, begin to limit the thermal conductivity. The scattering from the edges of graphene ribbons also deserves more rigorous treatment due to the large MFP in 2D graphene.

In order to study the thermal conductivity dependence on the lateral size of the graphene ribbon we consider the phonon anharmonic interactions to the second order and the angle dependence of the phonon scattering from the ribbon edges. We specifically focus on ribbons with the micrometer width $d$ and length $L$ in order to deal with the actual phonon dispersion in graphene and to ensure the diffusive transport regime. In the nanometer-thick graphene ribbons the phonon dispersion is different owing to the phonon mode quantization and the lateral size dependence is dictated by the ballistic conduction [32]. The total phonon scattering rate for the phonon mode ($s$, $q$) is given as

$$1/\tau_{tot,s}(\vec{q}) = 1/\tau_{U,s}(\vec{q}) + 1/\tau_{2,s}(\vec{q}) + 1/\tau_{B,s}(\vec{q}), \qquad (4)$$

where $\tau_{U,s}$ is the phonon mode-dependent three-phonon Umklapp scattering rate calculated to the first order, $\tau_{2,s}$ is the mode-dependent three-phonon scattering rate calculated to the second order [13,41] and $\tau_{B,s} = \Lambda_b(s,\vec{q},p)/v_s^{\parallel}(\vec{q})$ is the phonon mode-dependent boundary scattering rate, where $p$ is the specularity parameter. We perform the calculation of $\tau_{U,s}$ using our diagram technique described in details elsewhere [25-26]. The phonon mode-dependent MFP $\Lambda_b(s,\vec{q},p)$ limited by the boundary scattering is calculated as a function of the angle $\varphi$ between $\vec{q}$ and the thermal gradient for each phonon mode. Therefore in the case of rectangular ribbon $\tau_{B,s}$ depends both on $L$ and $d$ (see Figure 1 (b)).

In order to evaluate $\tau_{2,s}$ we include the following processes: the long-wavelength phonon $\vec{q}$ interacts with the short-wavelength phonon $\vec{q}\,'$ in the normal process forming a phonon $\vec{q}_i$. The phonon $\vec{q}_i$ then interacts with the phonon $\vec{q}\,''$ in the Umklapp process forming a phonon $\vec{q}\,'''$. The scattering rate of such processes in graphene takes the form

$$\frac{1}{\tau_{2,s}} = \frac{32}{9}\left(\frac{k_B T}{M(v_s^{\parallel})^2}\right)^2 \gamma_s^4 \int \left(\frac{a}{2\pi}\right)^4 (\omega')^2 \delta(\Delta\omega)d\vec{q}\,'d\vec{q}\,'' \qquad (5)$$

We derived Eq. (5) following the approach described in Ref. [41] and taking into account 2D phonon density of states in graphene. Considering all possible three-phonon processes in graphene using a formalism, derived by us in Ref. [26], we found that in the normal processes with the long-wavelength phonons $|q|<0.05q_{max}$, intensively participate phonons with $|\vec{q}\,'|\sim(0.6\text{-}0.7)q_{max}$, forming the phonons $|q_i|\sim(0.55\text{-}0.75)q_{max}$, while in the Umklapp processes, the phonons $|q_i|$ intensively interact with the phonons $|\vec{q}\,''|\sim(0.5\text{-}0.7)q_{max}$. Therefore, we can assume that for the most intensive second-order processes $|\vec{q}\,'|$ is close to $(0.6\text{-}0.7)q_{max}$, $|\vec{q}\,''|$ is close to $(0.5\text{-}0.7)q_{max}$, and can rewrite Eq. (5) as follows

$$\frac{1}{\tau_{2,s}} = \frac{2\pi}{9} \left( \frac{k_B T}{M(v_s^{\parallel})^2} \right)^2 \gamma_s^4 \omega_{\max,s}. \tag{6}$$

A similar formula was derived by Mingo and Broido [13] for CNTs. The thermal conductivity $K$ of graphene flakes was calculated using Eq. (1), substituting $\tau_{tot,s}(\vec{q})$ from Eq. (4) instead of $\tau_{U,s}$ and taking into account the actual graphene phonon energy spectrum determined from VFF method [26].

Figure 5 shows the dependence of the RT phonon thermal conductivity of the rectangular graphene ribbon on the ribbon length $L$ for different specular parameters $p$ and the ribbon width $d$. The specularity parameter $0<p<1$ determines the fraction of the diffusively scattered phonons contributing to the thermal resistance and is defined by the edge roughness [42]. The long-wavelength phonons weakly participate in three-phonon Umklapp processes. Therefore, their contribution to the thermal conductivity is mostly limited by the boundary scattering up to the length scale $L \sim 100$ μm. For $L>100$ μm the second order anharmonic processes become the main scattering mechanism for the long wavelength phonons. The most striking feature in Figure 5 is a non-monotonic dependence of the thermal conductivity on the ribbon length $L$. Such an unusual $K(L)$ characteristic suggests that the measured thermal conductivity of graphene ribbons of certain length, i.e. $L/d$ ratios, will be higher than that of graphene samples of other sizes and geometries.

[Figure 5]

We explain the possibility of the non-monotonic dependence via the following considerations. A portion of the acoustic phonons in the rectangular ribbon with the angle $\varphi < \arcsin(d/\sqrt{d^2 + L^2})$ does not scatter from the ribbon edges. MFP of these phonons $\Lambda_b = L/\cos(\varphi)$ is determined only by the ribbon length $L$ (at fixed $d$) and schematically shown in Figure 1 (b) by the violet and pink arrows. The rest of the phonons participate in the edge scattering and their $\Lambda_b$ depends on both $L$ and $d$ (schematically shown in Figure 1 (b) by the blue and green arrows): $\Lambda_b \approx \sqrt{(d \cdot n)^2 + L^2}$ if $n \leq (1+p)/(1-p)$ and $\Lambda_b = d \cdot (1+p)/(1-p)$ otherwise, where $n$ shows a number of reflections from the ribbon boundary. We calculated the number of reflections $n$ numerically (at fixed $L$, $d$, $\varphi$) from the condition $\Lambda_b \cdot \cos(\varphi) \leq L$. The interplay between contributions of the above-mentioned two groups of phonons as well as the anisotropic anharmonic scattering mechanisms leads to the predicted non-monotonic behavior of the thermal conductivity $K(L)$.

At small $L$ the phonons with the MFP limited by the length only – $\Lambda_b(L)$ – are the main heat carriers and thermal conductivity rapidly increases with $L$. The contribution of these type of phonons to the thermal conductivity in graphene ribbon with $d = 1$ μm is shown in Figure 5 (b) with the dashed line. Further increase of $L$ decreases $\alpha$ with the corresponding reduction of the number of phonons with $\Lambda_b(L)$ and increase of the number of phonons that have MFP dependent on both $L$, $d$ and $p$ – $\Lambda_b(L,d,p)$. Therefore, the contribution of the phonons with $\Lambda_b(L,d,p)$ increases (as shown in Figure 5 (b) with dotted line) leading to a maximum in the thermal conductivity curve. For $L >$~100 μm $\Lambda_b$ is mainly determined by $d$ and the thermal conductivity saturates to its finite value. The finite value in Figure 5 (b) for $d=5$ μm is in agreement with the experimental data [19-22]. The values for ribbons with large $d$ and $p\rightarrow 1$ are larger than what was reported experimentally because our model intentionally does not include non-idealities such as defects or grain boundaries.

Another important observation from Figure 5 (a) is that the abnormal non-monotonic $K(L)$ dependence can only be observed in graphene ribbons with the relatively smooth edges characterized by the specularity parameter $p>0.5$. The specularity parameter $p=1$ means that all phonons scatter from the edges elastically preserving their momentum along the ribbon length. Such scattering events do not contribute to the thermal resistance of the sample. The graphene ribbons with smooth edges are feasible technologically via a number of different techniques, e.g. unzipped CNTs or mechanical exfoliation [43-47].

The suspended graphene ribbons of rectangular shape and high-quality edges have also been demonstrated [19-22]. Thus, the $p>0.5$ requirement is not too restrictive for observation of the non-monotonic *K(L)* experimentally.

The *K(L)* non-monotonic dependence is also a function of the specific geometry of the ribbon via the angle $\varphi$ dependence on *L* and *d*. The non-monotonic character disappears in circular geometry such as in membranes used in some of the graphene thermal experiments [22, 48]. The study of the radius-dependence of the thermal conductivity in CNTs demonstrated the monotonic increase of *K* with the radius *R* until the constant values is reached at $R \sim 10 - 100$ μm [13]. Our results for the infinitely-wide ribbons ($d \to \infty$) also show the monotonic increase of *K* with the saturated value for *L*>100 μm. This finding is in line with the predictions made for the CNTs [13]. As in the case for CNTs, the thermal conductivity of graphene ribbons limited only by three-phonon Umklapp scattering increases monotonically with *L* (see dashed-dotted curve in Figure 5(b)) without saturation to the constant value. The finite value results from inclusion of the anharmonic three-phonon processes of the second-order.

The present study differs from previous reports of the size dependence of the thermal conductivity of graphene ribbons [26-32, 35, 49]. We consider the micrometer size ribbons and take into account both the ribbons length and size. In the case of infinitely wide ribbons we obtain the "conventional" monotonic dependence of the thermal conductivity on *L* in agreement with previous reports. The saturated values of the infinitely long ribbons are in agreement with those in Ref. [26]. The anisotropic phonon anharmonic scattering [49] in combination with the angle-dependent boundary scattering result in the unusual non-monotonic dependence of the thermal conductivity on *L*. One should note here that Haskins et al. [31] predicted a weaker non-monotonic dependence of the thermal conductivity on the width for 100-nm long zigzag nanoribbons. Although their result cannot be compared directly with our data for the micrometer-size ribbons it provides an example of another situation where the thermal conductivity is strongly affected by the shape and edge scattering of the graphene sample. The experimental study of the size dependence of the thermal conductivity in graphene ribbons can be performed using the optothermal Raman technique or electrical self-heating method [1, 19-25, 50]. However such a study is challenging owing to difficulties of preparation of defect-free graphene ribbons with different lateral dimensions and the same shape.

In conclusion, we investigated the thermal conductivity *K* of graphene ribbons and graphite slabs as the function of their lateral dimensions. Our results suggest that the long mean free path of the long-wavelength acoustic phonons in graphene results in abnormal *non-monotonic* dependence of the thermal conductivity on the length and width of a ribbon. Moreover, our analytical derivations also indicate that the bulk-like 3D phonons in graphite make a rather substantial contribution to its in-plane thermal conductivity. The predicted non-monotonic dependence of the thermal conductivity on the length of the ribbon for the micrometer-size graphene samples can account for some of the discrepancy in reported experimental data for graphene.

**Acknowledgments**

The authors acknowledge illuminating discussion with the late Professor E.P. Pokatilov who was involved in the early stage of this project. The work at UCR was supported, in part, by the National Science Foundation (NSF) projects US EECS-1128304, EECS-1124733 and EECS-1102074, by the US Office of Naval Research (ONR) through award N00014-10-1-0224, Semiconductor Research Corporation (SRC) and Defense Advanced Research Project Agency (DARPA) through FCRP Center on Functional Engineered Nano Architectonics (FENA), and DARPA-DMEA under agreement H94003-10-2-1003. DLN and ASA acknowledge financial support from the Moldova State Project No. 11.817.05.10F and Project for Young Scientists No. 12.819.05.18F.

**Figure Captions**

**Figure 1:** (a) Schematics of a typical suspended graphene ribbon used for experimental studies of thermal transport in suspended graphene ribbons. (b) Graphene ribbon and notations used in the present model for accounting the angle-dependent phonon scattering from the ribbon edges.

**Figure 2:** Phonon energy spectrum in bulk graphite calculated using VFF method. Note that LO, TO, LA, TA, ZA phonon branches are nearly double-degenerate except for a region near the $\Gamma$ point.

**Figure 3**: Equal energy surfaces for LA (a) and TA (b) phonon polarization branches in graphite. The EES model representation is used to elucidate the role of 3D bulk phonons in heat conduction along graphite basal planes and obtain analytical expression for the thermal conductivity scaling with the lateral size of graphite slab.

**Figure 4:** Dependence of the in-plane thermal conductivity of graphite on the length $L$ of the graphite slab. Note that the Umklapp-limited thermal conductivity of 3D graphite reveals the size dependence even for rather long slabs near RT.

**Figure 5:** (a) Dependence of the thermal conductivity of the rectangular graphene ribbon on the ribbon length $L$ shown for different specular parameters $p$. The width is fixed at $d=5$ μm. (b) Dependence of the thermal conductivity of the rectangular graphene ribbon on the ribbon length $L$ shown for different ribbon width $d$. The specular parameter is fixed at $p=0.9$. Note in both panels an unusual non-monotonic length dependence of the thermal conductivity, which results from the exceptionally long MFP of the low-energy phonons and their angle-dependent scattering from the ribbon edges.

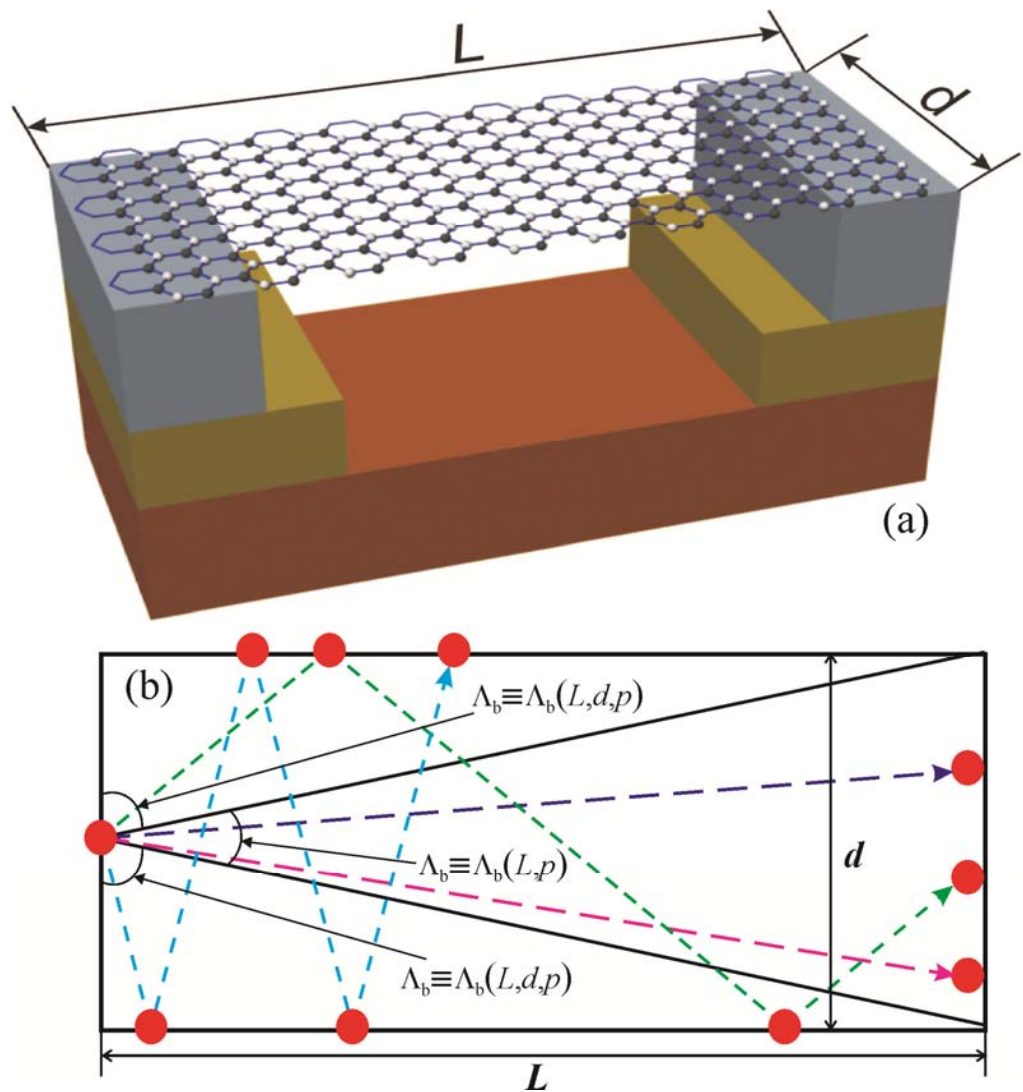

Figure 1 of 5: D.L. Nika, A.S. Askerov and A.A. Balandin

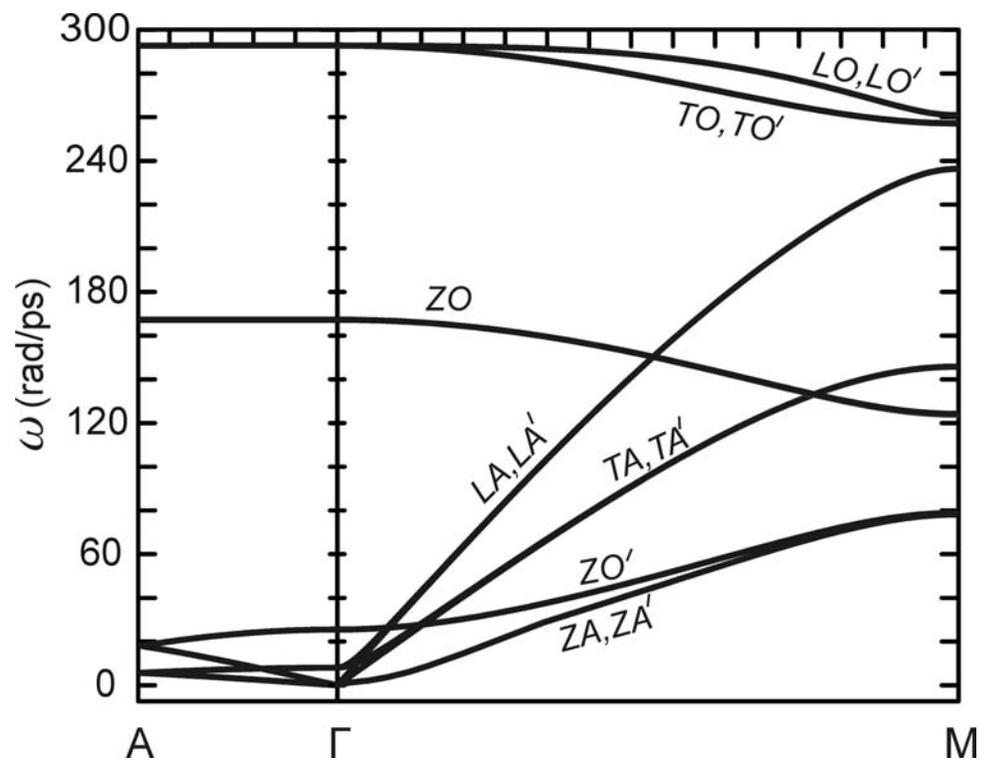

Figure 2 of 5: D.L. Nika, A.S. Askerov and A.A. Balandin

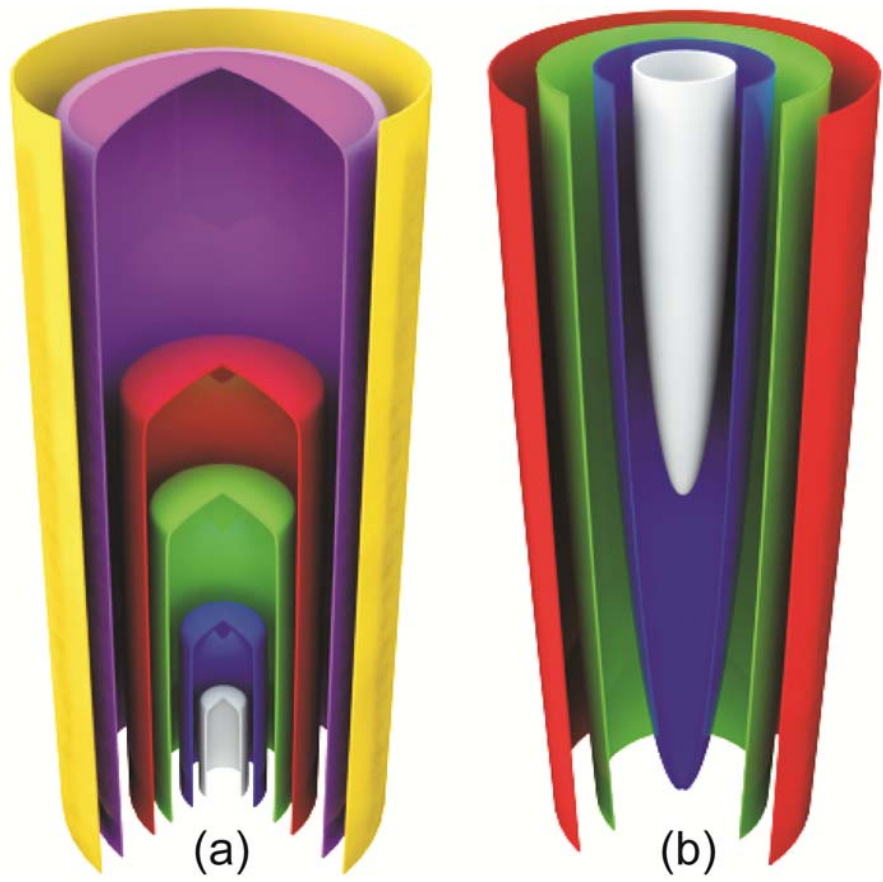

Figure 3 of 5: D.L. Nika, A.S. Askerov and A.A. Balandin

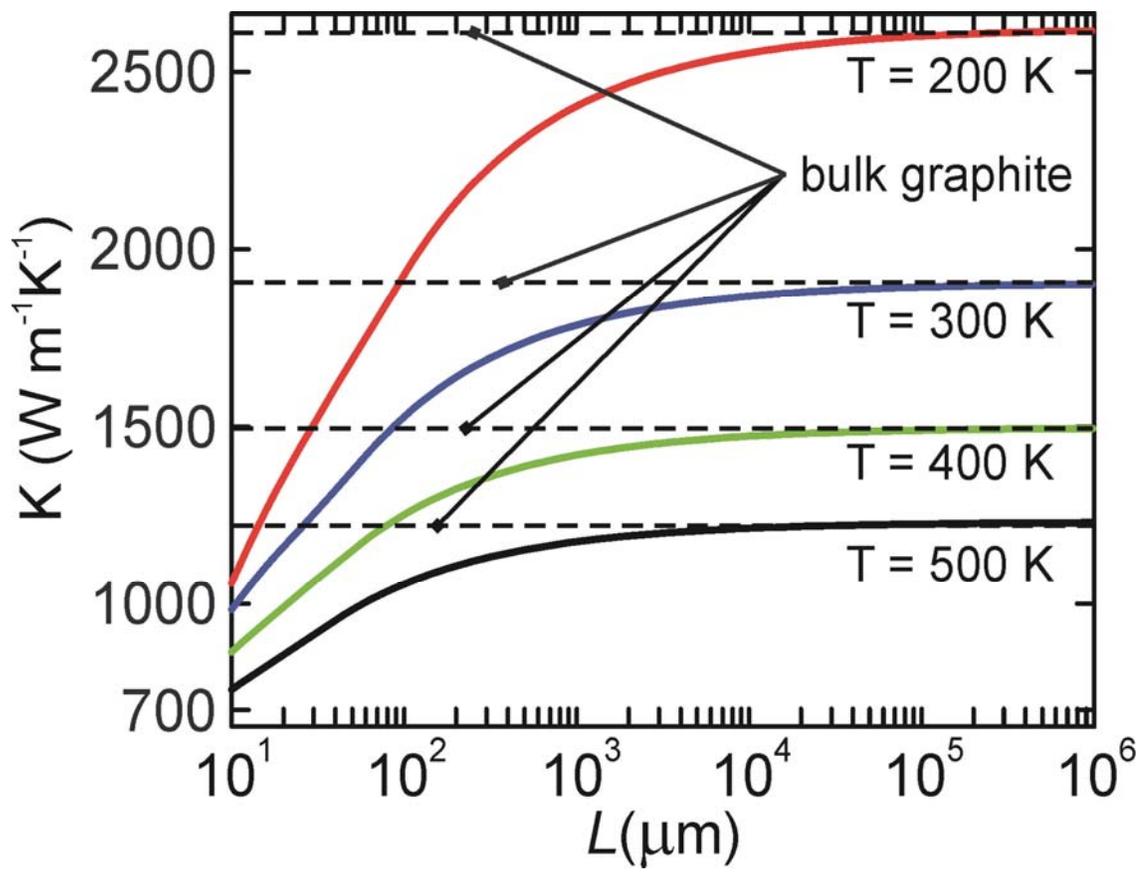

Figure 4 of 5: D.L. Nika, A.S. Askerov and A.A. Balandin.

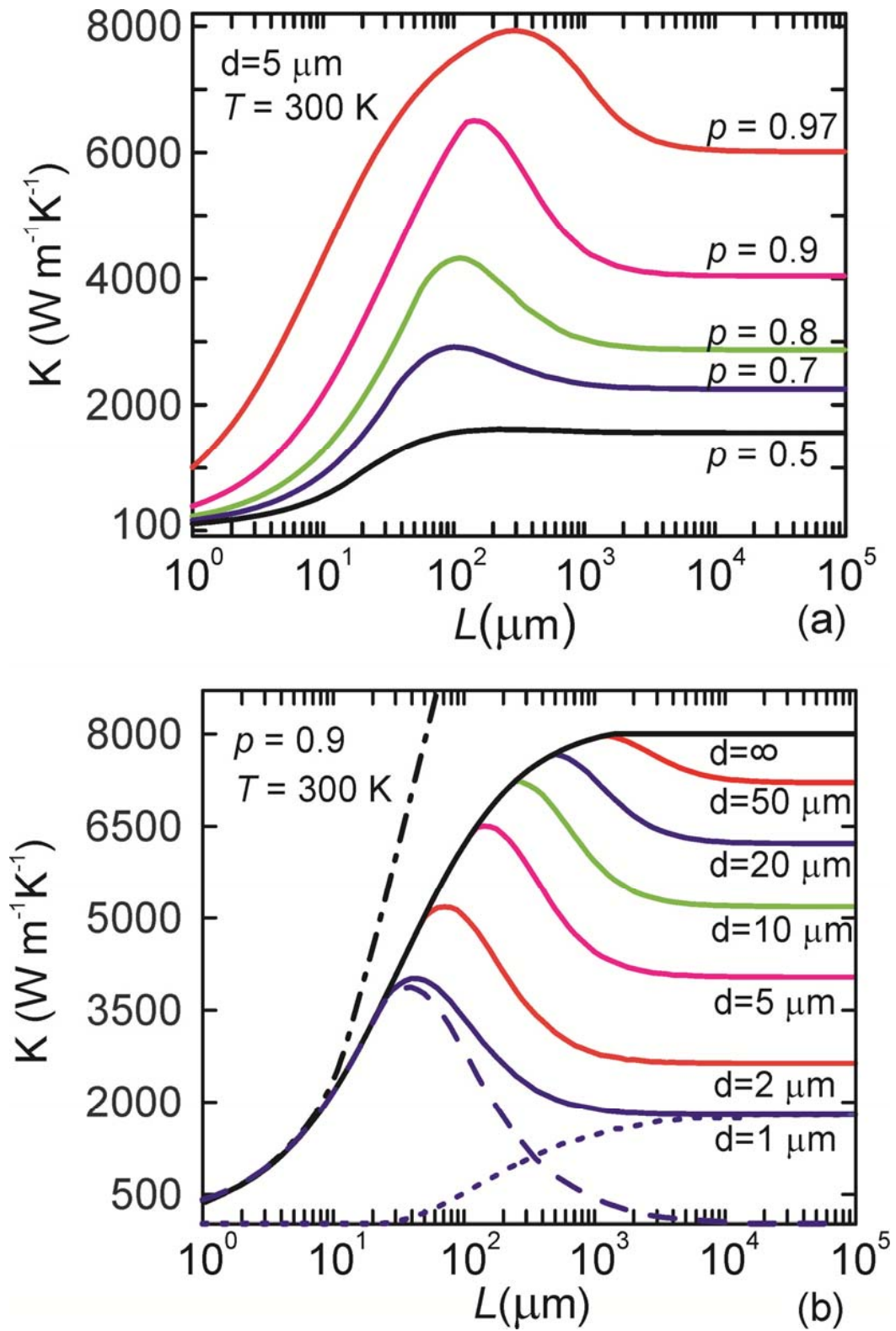

Figure 5 of 5: D.L. Nika, A.S. Askerov and A.A. Balandin